%% file: main2.tex
\documentclass[%
aps,pra,twocolumn,showpacs,superscriptaddress,floatfix
]{revtex4-1}
\pdfoutput=1

\usepackage[dvipdfmx,dvipdfmx]{graphicx}
\usepackage{graphicx}
\usepackage{color}
\usepackage{siunitx}

\usepackage{subfigure}

\usepackage{epsfig}
\usepackage{float}
\usepackage{epstopdf}
\usepackage{url}

\usepackage{amsmath}
\usepackage{amsfonts}

\newcommand{\ket}[1]{\left | #1 \right \rangle}
\newcommand{\bra}[1]{\left \langle #1 \right |}

\newcommand{\outerp}[2]{\left| #1 \left\rangle \right\langle #2 \right|}

\DeclareMathOperator{\tr}{Tr}

\def\be{\begin{equation}}
\def\ee{\end{equation}}
\usepackage{enumerate}

\newcommand{\affA}{%
\affiliation{
 Center for Macroscopic Quantum States (bigQ), Department of Physics, Technical University of Denmark, Building 307, Fysikvej, 2800 Kgs.~Lyngby, Denmark}
     }
\newcommand{\affB}{%
\affiliation{
National Institute of Information and Communications Technology (NICT), 4-2-1 Nukui-kita, Koganei, Tokyo 184-8795, Japan}}

\begin{document}

\title{Limit of Gaussian operations and measurements for Gaussian state discrimination, and its application to state comparison}

\author{David E.~Roberson}
\affA
\author{Shuro Izumi}
\affA
\author{Wojciech Roga}
\affB
\author{Jonas S. Neergaard-Nielsen}
\affA
\author{Masahiro Takeoka}
\affB
\author{Ulrik L. Andersen}
\affA

\begin{abstract}
{We determine the optimal method of discriminating and comparing quantum states from a certain class of multimode Gaussian states and their mixtures when arbitrary global Gaussian operations and general Gaussian measurements are allowed. We consider the so-called constant-$\hat{p}$ displaced states which include mixtures of multimode coherent states arbitrarily displaced along a common axis. 
We first show that no global or local Gaussian transformations or generalized Gaussian measurements can lead to a better discrimination method than simple homodyne measurements applied to each mode separately and classical postprocessing of the results. This result is applied to binary state comparison problems. 
We show that homodyne measurements, separately performed on each mode, are the best Gaussian measurement for binary state comparison.
We further compare the performance of the optimal Gaussian strategy for binary coherent states comparison with these of non-Gaussian strategies using photon detections.
}
\end{abstract}
\maketitle

\section{Introduction}\label{Sect:1}
Quantum state discrimination is the task of determining which quantum state, among a known set of states, a given system is in. The problem is non-trivial if the states are at least partially indistinguishable, i.e.~non orthogonal to each other \cite{Barnett2009}. The non-orthogonality arises not only due to imperfections of the measuring method or errors induced by limited knowledge and control but also due to the fundamental features of quantum mechanics.
This fundamental indistinguishability is the key enabling feature of quantum key distribution protocols \cite{bb84,b92,Gisin02,Lo14,Sasaki11} as it prevents an eavesdropper in extracting information from a quantum state without being noticed. 
However, the same features may imply restrictions when the results are to be read out \cite{lloyd08,tan08,pirandola11,pirandola11b,wilde12,roga15}, thereby limiting the mutual information between sender and recipient. To attain the maximum mutual information in quantum channels---the classical-quantum capacity---it is thus critical to optimize the discrimination scheme  
\cite{Holevo1998,Schumacher1997,Giovannetti2014}.

In quantum optical systems,
a natural set of resources are {\it Gaussian states, operations and measurements} \cite{braunstein05,cerf07,weedbrook12,adesso14}. 
Gaussian states are described by Gaussian characteristic functions on the phase space of the quadratures while 
Gaussian operations by definition preserve the Gaussianity of the characteristic functions. 
Generalized Gaussian measurements can be thought of as any Gaussian operation, partial Gaussian measurement and classical feedforward/feedback, followed by heterodyne measurements \cite{Takeoka2008}.
The technology of Gaussian operations and measurements are nowadays relatively well established and easily implementable but this limited set of transformation is insufficient for many quantum information protocols.
For example, by exploiting pure Gaussian transformation, quantum computation cannot show a quantum advantage \cite{Barlett02}, entanglement cannot be distilled \cite{Eisert02}, quantum error correction against Gaussian noise cannot be realized \cite{Niset08,Gagatsos19}, and the capacity of optical communication cannot be reached~\cite{PhysRevA.89.042309,PhysRevA.93.050302}.

Similarly, it has been shown that the optimal discrimination of binary phase shift keyed (BPSK) coherent states ($\ket{\alpha}, \ket{-\alpha}$) and thereby reaching the fundamental Helstrom bound, cannot be done by Gaussian measurements~\cite{helstrom}. To beat the Gaussian limit and approach the Helstrom bound, non-Gaussian measurements relying on photon detection have been theoretically conceived \cite{Kennedy73,Dolinar73,Bondurant1993,Guha2011,izumi2012,izumi2013} and experimentally realized \cite{CookMartinGeremia2007_Nature,Wittmann2008_PRL_BPSK,Tsujino2011_Q_Receiver_BPSK,Becerra13,Becerra15,PhysRevApplied.13.054015}. However, despite being insufficient for reaching the Helstrom bound, it is still interesting to find the optimal Gaussian approach that minimizes the error rate due to the simplicity of Gaussian measurements and their compatibility with current coherent communication systems.    
Indeed, it has been shown that among all possible Gaussian strategies, the optimal Gaussian strategy is simply to perform homodyne detection~\cite{Takeoka2008}.
However, for some important sets of Gaussian states or some particular noisy environments \cite{Olivares_2003,5407627}, the ultimate limit of the Gaussian schemes has not been fully investigated, and clarification of a Gaussian benchmark is of both practical and fundamental interest due to the simplicity in implementing Gaussian measurements. 


In this work we extend the results on the ultimate Gaussian limit in state discrimination to a much larger class of states. In particular, 
we consider the discrimination of any two mixtures of Gaussian states (squeezed thermal states) distributed along a certain line in phase space which we refer to as a \emph{constant-$\hat{p}$ set}. This includes, but is not restricted to, multimode coherent states displaced along a common axis. 
We show that for such states, the optimal Gaussian strategy is simply to perform homodyne detection on each mode which means there is no need for Gaussian multi-mode interactions, squeezing operations or feedback to attain the optimal Gaussian discrimination measurement.  
We also discuss the relation of our result to the task of quantum state comparison.
The goal of quantum state comparison is to assess if two states from a given set are the same or different. 
Indeed, in the case of binary state comparison, the optimal general strategy minimizing the error probability of the comparison is to simply perform the optimal discrimination measurement on each system and compare the outcomes ~\cite{barnett2003,PhysRevA.72.032308,statecompare}.
Therefore, as with state discrimination, one could expect that non-Gaussian measurements provide an advantage over Gaussian strategies for quantum state comparison. However, there is yet no rigorous benchmark for the ultimate performance of the Gaussian strategy for state comparison.
Here we show that homodyne measurements, individually performed on each system, is the best Gaussian strategy for minimizing the error probability for quantum state comparison. As for quantum state discrimination, there is no need for multimode interaction or classical feedforward/feedback to reach the optimal bound.

The paper is organized as follows. In Sec.~\ref{sec:gdiscriminate} we present our main result. We recognize the ultimate limit of fully Gaussian protocols for binary state discrimination of two arbitrary mixtures of constant-${\hat p}$ multimode states.   
We apply this result to several discrimination and comparison tasks in Sec.~\ref{sec:gaussstatecompare}. In Sec.~\ref{sec:nongaussian} we discuss as well practical non-Gaussian methods with
photon detections and the possibility to approach the theoretical bound for coherent state comparison.

\section{Gaussian binary state discrimination}\label{sec:gdiscriminate}

In binary state discrimination, we are provided a system prepared in one of two known states $\rho_1, \rho_2$ which may be mixed states in general. It is also assumed that we know the prior probability $p$ with which we receive $\rho_1$ (thus we receive $\rho_2$ with probability $1-p$). Our goal is to decide which state the given system is in with the largest possible success probability. If arbitrary measurements are allowed, then the optimal strategy is to perform a projective measurement whose two outcomes correspond to the positive and negative eigenspaces of the operator $p\rho_1 - (1-p)\rho_2$ which succeeds with probability
\begin{equation}
\frac{1}{2}\left(1 + \|p\rho_1 - (1-p)\rho_2\|_1\right),
\end{equation}
where $\|\rho\|_1$ is the sum of singular values of $\rho$ ~\cite{helstrom}.

\begin{figure}[t]
\centering
{
\includegraphics[width=0.95\linewidth]
{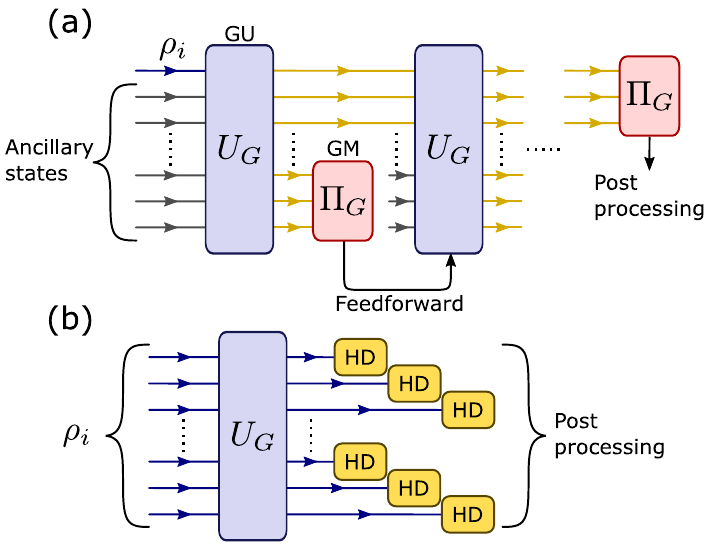}
}
\caption{
(a) General Gaussian strategy with classical feedforward operations.
GU:Gaussian unitary, GM:Gaussian measurement.
(b) 
The general gaussian strategy can be simplified to a measurement structure composed of Gaussian unitary operation followed by heterodyne detections (HD) performed on each mode.
}
\label{fig:stratgauss}
\end{figure}
\subsection{Gaussian measurement and state}\label{sec:statemeasurement}

We are interested in the optimal strategy when restricted to Gaussian operations and measurements, in the case where the two states are mixtures of Gaussian states. In Fig.~\ref{fig:stratgauss}(a), there is a schematic of a generic Gaussian operation. It consists of the $n$-mode input $\rho_i$, an $m$-mode ancillary state, a sequence of Gaussian unitary operations $U_G$ each followed by partial Gaussian measurements $\Pi_G$ whose outcomes are allowed to be fed forward, and finally a Gaussian measurement (which without loss of generality can be assumed to be a heterodyne measurement on each mode) followed by post-processing. The complicated nature of this protocol makes it difficult to analyze directly. However, it was shown that if the input is a mixture of Gaussian states, then the partial measurements and feed forward are unnecessary \cite{PhysRevA.89.042309}. Therefore, we may assume that our strategy consists of performing a single Gaussian unitary operation $U_G$ on our $n$-mode input state followed by a heterodyne measurement on each mode and then post-processing, as represented in Fig.~\ref{fig:stratgauss}(b). The outcome of such a measurement is given by a vector of $n$ complex numbers $\vec{\alpha} = (\alpha_1, \ldots, \alpha_n)$ corresponding to the $n$ heterodyne measurements. This outcome corresponds to POVM operator
\begin{equation}
 \Pi_{\vec{\alpha}} = \frac{1}{\pi^n}\outerp{\alpha_1}{\alpha_1} \otimes \ldots \otimes \outerp{\alpha_n}{\alpha_n}.
 \end{equation}
For discrimination of states $\rho_1$ and $\rho_2$, the post-processing will consist of partitioning this uncountable set of possible outcomes into two outcomes corresponding to whether we decide the received state is $\rho_1$ or $\rho_2$. This will result in POVM operators
\begin{eqnarray}
    \Pi'_1 &=& \int_{R_1} d\vec{\alpha} \Pi_{\vec{\alpha}}, \\
    \Pi'_2 &=& \int_{R_2} d\vec{\alpha} \Pi_{\vec{\alpha}},
\end{eqnarray}
where $R_1$ and $R_2$ partition $\mathbb{C}^n$. We can also incorporate the Gaussian unitary operation $U_G$ directly into the measurement, which will result in POVM elements
\begin{eqnarray}
    \Pi_1 &=& \int_{R_1} d\vec{\alpha} U_G\Pi_{\vec{\alpha}}U_G^\dagger, \\
    \Pi_2 &=& \int_{R_2} d\vec{\alpha} U_G\Pi_{\vec{\alpha}}U_G^\dagger.
\end{eqnarray}
Note that a Gaussian unitary by definition maps Gaussian states to Gaussian states. Therefore, the operator
\begin{equation}
U_G\Pi_{\vec{\alpha}}U_G^\dagger = \frac{1}{\pi^n}U_G\left(\outerp{\alpha_1}{\alpha_1} \otimes \ldots \otimes \outerp{\alpha_n}{\alpha_n}\right)U_G^\dagger,
\end{equation}
is a scalar multiple of a projection onto a Gaussian state, and thus the POVM elements $\Pi_1$ and $\Pi_2$ are integrals over Gaussian states. This implies that the Wigner function of these POVM elements is positive, which we will make use of later.

Now consider the case where $\rho_1$ and $\rho_2$ are mixtures of $n$-mode Gaussian states, i.e.,
\begin{equation}\label{eq:mixture}
    \rho_i = \sum_{j=1}^{m} p_i^j \tau_j
\end{equation}
where $p_i^j \ge 0$, and $\sum_{j=1}^{m} p_i^j = 1$ for $i = 1,2$ and $\tau_j$ is an $n$-mode Gaussian state for $j = 1, \ldots, m$ for some $m$. Note that using the same set of states for $\rho_1$ and $\rho_2$ is not a restriction, since we allow $p_i^j = 0$. We also remark that we could consider mixtures defined in terms of integrals over Gaussian states weighted by a probability density and the analysis would remain the same, but the application to state comparison 
makes finite mixtures more relevant for our work.
Our analysis does not hold for arbitrary mixtures of Gaussian states $\rho_1$ and $\rho_2$; we must put some restrictions on the Gaussian states making up these mixtures. To describe these restrictions, we briefly review the basics of Gaussian states.

Recall that a Gaussian state is completely described by its first and second moments of the quadrature operators, i.e., its displacement vector $d$ and covariance matrix $\Gamma$. For an $n$-mode state, the displacement vector is a $2n$-dimensional real vector and the covariance matrix is a $2n \times 2n$ real symmetric positive definite matrix. The entries of $d$ and the rows/columns of $\Gamma$ are indexed by the quadrature
operators for each mode, usually in the order $\hat{x}_1, \hat{p}_1, \ldots, \hat{x}_n, \hat{p}_n$. However, it is more convenient for us to index them in the order $\hat{x}_1, \ldots, \hat{x}_n, \hat{p}_1, \ldots, \hat{p}_n$, which we will do from here on. The quadrature operators satisfy the commutation relations $[\hat{x}_\ell,\hat{p}_k]=i\delta_{\ell k}$, where $\delta_{\ell k}$ is the Kronecker delta and we use the convention $\hbar=1$. Thus for a given Gaussian state we can write its covariance matrix and displacement vector as
\begin{equation}
\Gamma = \begin{pmatrix}\Gamma_x & \Gamma_{xp} \\ \Gamma_{xp}^T & \Gamma_p\end{pmatrix} \quad d = \begin{pmatrix}d_x \\ d_p\end{pmatrix}.
\end{equation}

Suppose now that $\rho_1$ and $\rho_2$ are mixtures of $n$-mode Gaussian states as written in Eq.~\eqref{eq:mixture}. For each $j = 1, \ldots, m$ let
\begin{equation}\label{covmats}
    \Gamma^{j} = \begin{pmatrix} \Gamma^{j}_x & \Gamma^{j}_{xp} \\ (\Gamma^{j}_{xp})^T & \Gamma^{j}_p\end{pmatrix} \text{ and } d^{j} = \begin{pmatrix}d^{j}_x \\ d^{j}_p\end{pmatrix},
\end{equation} 
be the covariance matrix and displacement vector of the state $\tau_j$. We consider the case where there exists a fixed $\Gamma_p$ and $d_p$ such that $\Gamma^{j}_p = \Gamma_p$, $d^{j}_p = d_p$, and $\Gamma^{j}_{xp} = 0$ for all $j = 1, \ldots, m$. We refer to such a set of Gaussian states as a \emph{constant-$\hat{p}$} set. Notice that $\tau_1$ can be a arbitrary multimode displaced squeezed thermal state with diagonal covariance matrix which determines $d_p$ and $\Gamma_p$ for the remaining states in the mixture Eq.(\ref{eq:mixture}). However, apart from the present section, we focus our attention on coherent states, so with zero noise and squeezing. We will show that if $\rho_1$ and $\rho_2$ are mixtures of Gaussian states from a constant-$\hat{p}$ set, then the optimal Gaussian strategy for discriminating $\rho_1$ and $\rho_2$ is to perform a homodyne measurement in the $\hat{x}$-quadrature on each mode.

Our analysis will make use of the Wigner function formalism \cite{wigner32,braunstein05} of quantum states and operators. For any $n$-mode linear operator $X$, its Wigner function is 
\begin{eqnarray}
    W_X(\vec{x},\vec{p}) &=& W_X(x_1, \ldots, x_n, p_1, \ldots p_n) 
    \nonumber
    \\
    &=& \int d^n \vec{u} \ e^{i\vec{u}\vec{p}} \bra{\vec{x}+\frac{\vec{u}}{2}}X\ket{\vec{x}-\frac{\vec{u}}{2}},
\end{eqnarray}
where $\ket{\vec{x}} = \ket{x_1} \otimes \ldots \otimes \ket{x_n}$ are the quadrature eigenstates. Two of the main properties of the Wigner function that we will use are that it is linear in $X$ and that for a Gaussian state $\rho$ with covariance matrix $\Gamma$ and displacement vector $d$ the Wigner function evaluates to
\begin{equation}\label{eq:gausswigner}
    W_\rho(\vec{r}) = \frac{1}{\pi^n\sqrt{\det(\Gamma)}}e^{-(\vec{r} - d)^T\Gamma^{-1}(\vec{r}-d)},
\end{equation}
where $\vec{r} = (\vec{x}, \vec{p})^T$. We also use the fact that the overlap of two linear operators $X$ and $Y$ can be written in terms of their Wigner functions:
\begin{equation}
\tr(XY) = \int d\vec{x} \ d\vec{p} \ W_X(\vec{x},\vec{p})W_Y(\vec{x},\vec{p}).
\end{equation}

\subsection{Optimal Gaussian measurement }\label{sec:optmeasurement}

Assuming, as above, that we are given $\rho_1$ with probability $p$, we can write the error probability of our Gaussian discrimination protocol as
\begin{eqnarray}
    P_\text{err} &=& p\tr(\Pi_2 \rho_1) + (1-p)\tr(\Pi_1 \rho_2) 
    \nonumber
    \\
    &=& p\tr(\Pi_2\rho_1) + (1-p)\tr((I - \Pi_2) \rho_2) 
    \nonumber
    \\
    &=& (1-p) + \tr\left[\Pi_2\left(p\rho_1 - (1-p)\rho_2\right)\right],
\end{eqnarray}
where we have used the fact that $\Pi_1 + \Pi_2 = I$. Let $X = p\rho_1 - (1-p)\rho_2$. In order to minimize the error probability, we must choose $\Pi_2$ such that $\tr(\Pi_2 X)$ is minimized. In terms of Wigner functions, we wish to minimize
\begin{equation}\label{eq:wignertrace}
    \int d\vec{x} \ d\vec{p} \ W_{\Pi_2}(\vec{x},\vec{p})W_X(\vec{x},\vec{p}).
\end{equation}

Now, recalling that $\Pi_2$ is an integral over Gaussian states, we have that $W_{\Pi_2}(\vec{x}, \vec{p}) \ge 0$ for all $\vec{x}, \vec{p}$. Moreover, since $\Pi_1 + \Pi_2 = I$, and the Wigner function of $I$ is 1 everywhere, we have that $W_{\Pi_2}(\vec{x}, \vec{p}) \le 1$ for all $\vec{x}, \vec{p}$. We can therefore lower bound the expression in~Eq.\eqref{eq:wignertrace} by
\begin{equation}\label{eq:wignerint}
    \int_R d\vec{x} \ d\vec{p} \ W_X(\vec{x}, \vec{p}),
\end{equation}
where $R$ is the region where $W_X(\vec{x}, \vec{p})$ is negative. Recall that the operator $X$ is a linear combination of Gaussian states $\tau_j$ for $j = 1, \ldots, m$. Consider the Wigner function for a single such state $\tau_j$. By~Eq.\eqref{eq:gausswigner} and our assumption on the covariance matrix and displacement vector of $\tau_j$, we have that

\begin{widetext}
\begin{eqnarray}
    W_{\tau_j}(\vec{x},\vec{p}) &=& \frac{\exp\left[-(\vec{x} - d^{j}_x)^T(\Gamma^{j}_x)^{-1}(\vec{x}-d_x^{j}) - (\vec{p} - d_p)^T\Gamma_p^{-1}(\vec{p}-d_p)\right]}{\pi^n\sqrt{\det(\Gamma_x^{j})\det(\Gamma_p)}}
    \nonumber
    \\
    &=& \left(\frac{\exp\left[-(\vec{x} - d_x^{j})^T (\Gamma_x^{j})^{-1}(\vec{x} - d_x^{j})\right]}{\sqrt{\pi^n \det(\Gamma_x^{j})}}\right)  \left(\frac{\exp\left[-(\vec{p} - d_p)^T \Gamma_p^{-1}(\vec{p} - d_p)\right]}{\sqrt{\pi^n \det(\Gamma_p)}}\right).
\end{eqnarray}
\end{widetext}

In other words, the Wigner function of $\tau_j$ factors as
\begin{equation}
    W_{\tau_j}(\vec{x},\vec{p}) = f_{j}(\vec{x})f(\vec{p}),
\end{equation}
where $f(\vec{p})$ does not depend on $j$. Furthermore, $f_{j}(\vec{x})$ and $f(\vec{p})$ are Gaussian probability distributions in $\vec{x}$ and $\vec{p}$ respectively. Therefore
\begin{eqnarray}
    f_{j}(\vec{x}) &=& \int_{\mathbb{R}^n}d\vec{p} \ W_{\tau_j}(\vec{x},\vec{p}), \\
    f(\vec{p}) &=& \int_{\mathbb{R}^n}d\vec{x} \ W_{\tau_j}(\vec{x},\vec{p}).
\end{eqnarray}
Thus $f_{j}(\vec{x})$ and $f(\vec{p})$ are the probability distributions over outcomes resulting from homodyning each mode of $\tau_j$ in the $\hat{x}$- and $\hat{p}$-quadratures respectively.

The Wigner function $W_X(\vec{x}, \vec{p})$ is a linear combination of the Wigner functions $W_{\tau_j}(\vec{x},\vec{p})$, and thus we can also factor the $f(\vec{p})$ term out of the former. Thus $W_X(\vec{x},\vec{p}) = g(\vec{x})f(\vec{p})$ where
\begin{equation}
    g(\vec{x}) = pg_1(\vec{x}) - (1-p)g_2(\vec{x}),
\end{equation}
and
\begin{equation}
    g_i(\vec{x}) = \sum_{j=1}^{m} p_i^j f_{j}(\vec{x})
    = \int_{\mathbb{R}^n} d\vec{p} \ W_{\rho_i}(\vec{x},\vec{p}),
\end{equation}
i.e.,~$g_i(\vec{x})$ is the probability distribution resulting from homodyning each mode of $\rho_i$ in the $\hat{x}$-quadrature.

Since $W_X(\vec{x},\vec{p}) = g(\vec{x})f(\vec{p})$ and $f(\vec{p})$ is positive everywhere (since it is a Gaussian probability distribution), the region $R \subseteq \mathbb{R}^{2n}$ where $W_X(\vec{x}, \vec{p})$ is negative only depends on $g(\vec{x})$. Letting $R_x \subseteq \mathbb{R}^n$ be the region where $g(\vec{x})$ is negative, we can write the expression in~Eq.\eqref{eq:wignerint} as
\begin{equation}
    \left(\int_{R_x} d\vec{x} \ g(\vec{x})\right)\left(\int_{\mathbb{R}^n} d\vec{p} \ f(\vec{p})\right) 
    = \int_{R_x} d\vec{x} \ g(\vec{x}),
\end{equation} 
where we have used the fact that $f(\vec{p})$ is a probability distribution. Plugging this into our lower bound on the error of our Gaussian discrimination protocol, we obtain
\begin{equation}\label{eq:lowerbd}
    P_\text{err} \ge (1-p) + \int_{R_x} d\vec{x} \ g(\vec{x}).
\end{equation}

Now let us consider a discrimination protocol for $\rho_1$ and $\rho_2$ which consists simply of performing homodyne detection in the $\hat{x}$-quadrature on each mode and then post-processing. We will see that with such a protocol we are able to obtain an error probability equal to the lower bound given in~Eq.\eqref{eq:lowerbd}, thus proving optimality.

After performing the $\hat{x}$-quadrature homodyne detection on each mode, we will obtain an outcome $\vec{x} \in \mathbb{R}^n$. Our post-processing procedure then takes this outcome and determines whether we should conclude that the state we were given was $\rho_1$ or $\rho_2$. Thus our post-processing can be specified by a subset $S \subseteq \mathbb{R}^n$ such that if our outcome $\vec{x} \in S$, then we conclude that we were given $\rho_2$ and otherwise conclude we were given $\rho_1$. As we have already seen, the probability distribution over outcomes resulting from $\hat{x}$-quadrature homodyne detection on each mode of $\rho_i$ is $g_i(\vec{x})$. Thus the error of this protocol is
\begin{eqnarray}
    &&p\int_{S} d\vec{x} \ g_1(\vec{x}) + (1-p)\int_{\mathbb{R}^n \setminus S} d\vec{x} \ g_2(\vec{x}) 
    \nonumber
    \\
    &=& (1-p) + \int_S d\vec{x} \ \left[pg_1(\vec{x}) - (1-p)g_2(\vec{x})\right] 
    \nonumber
    \\
    &=& (1-p) + \int_S d\vec{x} \ g(\vec{x})
\end{eqnarray}
Thus, the lower bound can be obtained by performing an $\hat{x}$-quadrature homodyne detection on each mode of the given state, and concluding it was $\rho_2$ if the outcome was in $R_x$ and concluding the state was $\rho_1$ otherwise.

We remark that determining the region $R_x$ may be difficult in practice, but this does not prevent one from implementing the above described optimal Gaussian discrimination protocol. Indeed, one does not need to precompute $R_x$ in order to implement this protocol. Rather, after performing the homodyne detection on each mode and obtaining outcome $\vec{x}$, one simply computes $g(\vec{x})$ for that outcome. If it is negative then conclude the state was $\rho_2$ and otherwise conclude it was $\rho_1$.

Unfortunately, we do not know how to derive a closed form expression for the error probability of the above protocol. In the case where $p = 1/2$, the error probability is closely related to the \emph{total variation distance} of the two distributions $g_1(\vec{x})$ and $g_2(\vec{x})$, denoted $\mathrm{TV}(g_1,g_2)$. This is defined as
\begin{eqnarray}
    \mathrm{TV}(g_1,g_2) &=& \frac{1}{2}\int_{\mathbb{R}^n} d\vec{x} \  |g_1(\vec{x}) - g_2(\vec{x})| 
    \nonumber
    \\
    &=& \int_{\hat{S}} d\vec{x} \  \left(g_1(\vec{x}) - g_2(\vec{x})\right),
\end{eqnarray}
where $\hat{S} \subseteq \mathbb{R}^n$ is the region where $g_1(\vec{x}) > g_2(\vec{x})$. Thus when $p = 1/2$, the error probability is $\tfrac{1}{2}\left[1 - \mathrm{TV}(g_1,g_2)\right]$.

\section{Gaussian State Comparison}\label{sec:gaussstatecompare}

Formally, the task of state comparison for a set $S = \{\tau_1, \ldots, \tau_m\}$ of known quantum states consists of being given two states $\rho_1, \rho_2 \in S$, and deciding whether $\rho_1 = \rho_2$. It is assumed that the deciding agent knows the probability $p_{ij}$ of receiving the ordered pair of states $(\tau_i, \tau_j)$, and their objective is to maximize the probability of correctly determining whether the two states they are given are equal. In general, this means that they perform a measurement $\mathcal{M} = \{\Pi_E, \Pi_{D}\}$ on the state $\rho_1 \otimes \rho_2$ with two possible outcomes corresponding to whether they claim that the two states are equal $(E)$ or different ($D$). The expected probability of making an error is then given by
\begin{eqnarray}
\label{eq:error}
P_{\text{err}} &=& \sum_{i \ne j} p_{ij}\tr\left[\Pi_{E}\left(\tau_i \otimes \tau_j\right)\right] \nonumber
\\
&&+ \sum_{i} p_{ii}\tr\left[\Pi_{D}\left(\tau_i \otimes \tau_i\right)\right] 
\nonumber
\\
&=& \tr\left[\Pi_{E}\left(\sum_{i \ne j} p_{ij} \tau_i \otimes \tau_j\right)\right] 
\nonumber
\\
&&+ \tr\left[\Pi_{D}\left(\sum_{i} p_{ii} \tau_i \otimes \tau_i\right)\right] \nonumber
\\
&=& \tr\left[\Pi_{E}\left( p_{D} \rho_{D}\right)\right] + \tr\left[\Pi_{D}\left(p_{E} \rho_{E}\right)\right]
\end{eqnarray}
where $p_{E} = \sum_{i} p_{ii}$ and $\rho_{E} =  (1/p_{E}) \sum_{i} p_{ii} \tau_i \otimes \tau_i$, and similarly for $p_{D}$ and $\rho_{D}$. Thus the state comparison problem reduces to the state discrimination problem for states $\rho_{E}$ and $\rho_{D}$ given with prior probabilities $p_{E}$ and $p_{D}$ respectively.

Suppose that the set $S$ of states we are comparing is a constant-$\hat{p}$ set. Then there exists a matrix $\Gamma_p$ and vector $d_p$ such that the covariance matrix and displacement vector of every state $\tau_i \in S$ has the form
\begin{equation}
    \Gamma^i = \begin{pmatrix}\Gamma^i_x & 0 \\ 0 & \Gamma_p\end{pmatrix} \quad d^i = \begin{pmatrix}d^i_x \\ d_p\end{pmatrix}
\end{equation}
Thus the state $\tau_i \otimes \tau_j$ has covariance matrix and displacement vector equal to the following:
\begin{equation}
    \begin{pmatrix}\Gamma^i_x \oplus \Gamma^j_x& 0 \\ 0 & \Gamma_p \oplus \Gamma_p \end{pmatrix} \quad \text{and} \quad \begin{pmatrix}d^i_x \oplus d^j_x \\ d_p \oplus d_p\end{pmatrix}.
\end{equation}
This means that the Gaussian states $\tau_i \otimes \tau_j$ for $i,j = 1, \ldots, m$ form a constant-$\hat{p}$ set as well. Since both $\rho_E$ and $\rho_{D}$ are mixtures of these states, the result of Sec.~\ref{sec:gdiscriminate} can be applied. Therefore, the optimal Gaussian state comparison protocol for a constant-$\hat{p}$ set of Gaussian states is to perform homodyne detection in the $\hat{x}$-quadrature on each mode. We remark that the set $S$ can actually be slightly more general: it can consist of mixtures of Gaussian states from some constant-$\hat{p}$ set $T$.

\subsection{Gaussian binary state comparison}\label{gbinstcom}

Let $T = \{\tau_1, \tau_2\}$ be a set of two states on which we wish to perform state comparison. Suppose also that the probability of receiving the ordered pair $(\tau_i,\tau_j)$ follows a product distribution: $i = 1$ with probability $q$ and $j = 1$ with probability $q$ independently. In this case,
\begin{eqnarray}
    p_E\rho_E &=& q^2\tau_1\otimes \tau_1 + (1-q)^2\tau_2 \otimes \tau_2 
    \\
    p_{D}\rho_{D} &=& q(1-q)\left(\tau_1 \otimes \tau_2 + \tau_2 \otimes \tau_1\right)
\end{eqnarray}
Letting $X = q\tau_1 - (1-q)\tau_2$, it is easy to see that $X \otimes X = p_E\rho_E - p_{D}\rho_{D}$. Using the expression for error in~Eq.\eqref{eq:error} and substituting $\Pi_{D} = I - \Pi_E$, we have
\begin{eqnarray}
    P_\text{err} &= p_E + \tr\left[\Pi_E\left(p_{D} \rho_{D} - p_E\rho_E\right)\right] \\
    &= p_E - \tr\left[\Pi_E(X \otimes X)\right]
\end{eqnarray}
This follows the analysis in Ref.~\cite{statecompare}. They further note that since $\Pi_E$ is a positive operator between 0 and $I$, choosing it to be the projection onto the positive eigenspace of $X \otimes X$ minimizes the error. Thus, if $\Pi$ is the projection onto the positive eigenspace of $X$, then the optimal choice of $\Pi_E$ is $\Pi \otimes \Pi + (I-\Pi)\otimes (I-\Pi)$, where $\{\Pi, I - \Pi\}$ is the optimal POVM for discriminating $\tau_1$ and $\tau_2$ with prior probabilities $q$ and $1-q$ respectively. Thus they conclude that the optimal state comparison procedure is to perform optimal state discrimination on each received state, and then conclude the states were equal if they get the same outcomes, and otherwise conclude they were different. If the optimal success probability for the state discrimination was $p$, then the optimal success probability for state comparison will be $p^2+(1-p)^2$.

We aim to extend the above result of Ref.~\cite{statecompare} to the Gaussian case. In this case $\tau_1$ and $\tau_2$ are (mixtures of) some $n$-mode Gaussian states. Almost all of the analysis above still holds, except that we cannot freely pick the POVM operator,
rather we are restricted to Gaussian operations and measurements. It is thus not immediately obvious that the optimal choice for $\Pi_E$ will have the same form as above. However, if $\tau_1$ and $\tau_2$ are mixtures of $n$-mode Gaussian states from a constant-$\hat{p}$ set, then we can apply our previous results. Thus in this case the optimal measurement is homodyne detection on each of the $2n$ modes. Note however this does not fully specify the POVM, since that also depends on the post-processing of the homodyne detection outcomes. To determine the optimal post-processing, let $g_1(\vec{x})$ and $g_2(\vec{x})$ be the probability distributions resulting from performing homodyne detection on each mode of $\tau_1$ and $\tau_2$ respectively. Then the distribution obtained from performing homodyne detection on every mode of $\tau_i \otimes \tau_j$ is the product distribution $g_i(\vec{x}^1)g_j(\vec{x}^2)$, where the superscripts indicate whether the vector variable refers to the first or last $n$-modes. If we let $g(\vec{x}) = qg_1(\vec{x}) - (1-q)g_2(\vec{x})$ and let $R \subseteq \mathbb{R}^{2n}$ be the region of outcomes for which we conclude that the states were the same, then our error is given by
\begin{equation}
    P_\text{err} = q^2 + (1-q)^2 - \int_{R} d\vec{x}^1 \ d\vec{x}^2 \ g(\vec{x}^1)g(\vec{x}^2)
\end{equation}
Obviously, the optimal choice for $R$ is the region where $g(\vec{x}^1)g(\vec{x}^2)$ is positive. If we let $R' \subseteq \mathbb{R}^{n}$ be the region where $g(\vec{x})$ is positive, then an optimal choice for $R$ is $(R' \times R') \cup (\overline{R'} \times \overline{R'})$, where $\overline{R'} = \mathbb{R}^n \setminus R'$. Note that this will include some points where $g(\vec{x}^1)g(\vec{x}^2) = 0$, but this will not change the error probability. The region $R' \times R'$ corresponds to performing optimal Gaussian discrimination on each received state and determining that they are both $\tau_1$, while $\overline{R'} \times \overline{R'}$ corresponds to determining both are $\tau_2$. Thus the optimal Gaussian state comparison protocol for $\{\tau_1,\tau_2\}$ is to perform optimal Gaussian state discrimination on each state and conclude they are equal if and only if the outcomes are the same.

In the case of the BPSK states $\{\ket{\alpha},\ket{-\alpha}\}$, we have $\Gamma_x = \Gamma_p = I$, $\Gamma_{xp} = 0$, for both states, and $d = (\pm \sqrt{2}\alpha, 0)^T$ respectively. Assuming uniform priors, the optimal Gaussian protocol has an error of $\tfrac{1}{2}\left[1 - \mathrm{erf}^2(\sqrt{2}\alpha)\right]$, and the optimal general strategy has error $\tfrac{1}{2}e^{-4\alpha^2}$ 
(see Fig.~\ref{Error}).


\subsection{Coherent state comparison with non-Gaussian measurements}\label{sec:nongaussian}
The fundamental lower bound of $\frac{1}{2}e^{-4\alpha^2}$ on the error probability for binary state comparison of the BPSK states is attainable by separately performing an optimal projective measurement for the state discrimination on each mode \cite{statecompare}.
For the problem of discriminating the BPSK states, non-Gaussian measurements based on photon detection provide a notable performance overcoming the Gaussian limit \cite{Kennedy73,Dolinar73,Takeoka2008}.
In this section, we will investigate the potential of
non-Gaussian measurements consisting of a displacement operation and a photon detection in coherent state comparison. This is a promising and practical non-Gaussian measurement beating the Gaussian limit in state discrimination \cite{Kennedy73,Tsujino2011_Q_Receiver_BPSK}.

\begin{figure}[t]
\centering
{
\includegraphics[width=0.95\linewidth]
{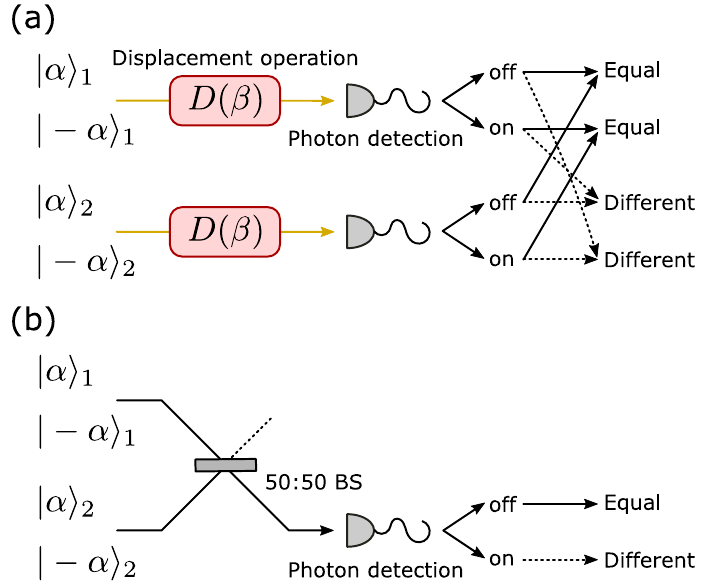}
}
\caption{
Schematics of coherent state comparison with,
(a) the displacement operation with the photon detection measurements individually performed on each mode,
(b) the balanced beam splitter with the photon detection.
}
\label{Non_gauss_scheme}
\end{figure}

A schematic of the coherent state comparison with such non-Gaussian measurement is shown in Fig.~\ref{Non_gauss_scheme}(a).
The displacement based photon detection measurements are individually performed on each mode, where one of the BPSK states is displaced close to a vacuum state.
We conclude whether the states are equal or different depending on the number of detector clicks, i.e., equal for even number of clicks and different for odd number of clicks.
The POVMs of the strategy using the displacement with photon detection measurements for the state comparison are given by
\begin{eqnarray}
\Pi_E^{NG} & = & \Pi_{\mathrm{off}}^{NG_1} \otimes \Pi_{\mathrm{off}}^{NG_2} + \Pi_{\mathrm{on}}^{NG_1} \otimes \Pi_{\mathrm{on}}^{NG_2},
\nonumber
\\
\Pi_{D}^{NG} & = & \Pi_{\mathrm{off}}^{NG_1} \otimes \Pi_{\mathrm{on}}^{NG_2} +\Pi_{\mathrm{on}}^{NG_1} \otimes \Pi_{\mathrm{off}}^{NG_2},
\end{eqnarray}
where the POVMs for the displacement plus photon detection measurement $\{\Pi_{\mathrm{off}}^{NG}, \Pi_{\mathrm{on}}^{NG} \}$ are represented by
\begin{eqnarray}
\Pi_{\mathrm{off}}^{NG} & = & D(\beta)^{\dagger} |0 \rangle \langle 0| D(\beta ),
\nonumber
\\
\Pi_{\mathrm{on}}^{NG} & = & I - \Pi_{\mathrm{off}}^{NG}.
\label{eq1}
\end{eqnarray}
If the displacement operation $D(\beta)$ is performed such that a state $\ket{\alpha}$ is displaced to a vacuum state,
the achievable error probability for the coherent state comparison is obtained to be
\begin{eqnarray}
P_\mathrm{err}^{NG} &=&\frac{1}{2}(\tr{[\rho_E\Pi_{D}^{NG}]}+ \tr{[\rho_{D}\Pi_E^{NG}]})
\nonumber
\\
&=&
e^{-4\alpha^2}(1-\frac{1}{2}e^{-4\alpha^2}).
\end{eqnarray}

\begin{figure}[t]
\centering
{
\includegraphics[width=0.95\linewidth]
{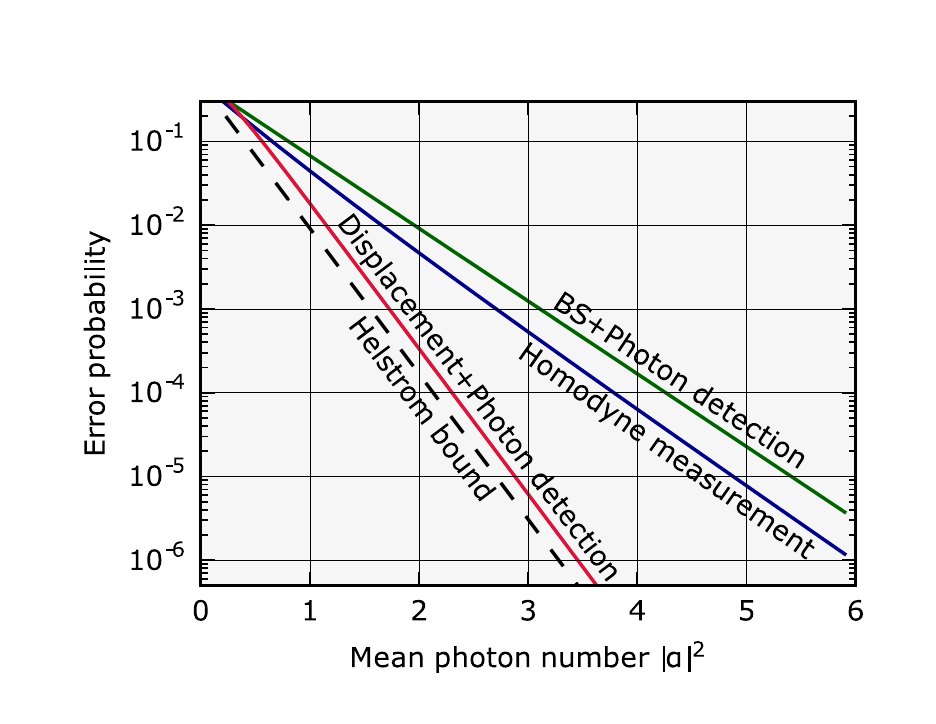}
}
\caption{
Error probability of coherent state comparison as a function of the signal mean photon number.
The black dashed line represents the Helstrom bound.
The red, blue and green solid lines are the performances of the strategies using displacement operation plus photon detection, homodyne measurement, and balanced beam splitter plus photon detection, respectively.
}
\label{Error}
\end{figure}

Figure \ref{Error} depicts the error probabilities of the coherent state comparison for various measurement strategies.
The Helstrom bound, shown by a black dashed line, is the fundamental bound of the discrimination error for a given pair of states.
A comparison scheme with the non-Gaussian measurement consisting of the displacement operation and the photon detection, shown by a red solid line, significantly outperform the Gaussian limit that is attainable by a homodyne measurement (blue solid line) and shows a near-optimal performance approaching the Helstrom bound.
Another non-Gaussian measurement for coherent state comparison can be implemented with a balanced beam splitter followed by a photon detection as shown in Fig.\ref{Non_gauss_scheme} (b).
While this strategy is technically simple because it does not require additional phase reference fields, and is an optimal measurement for unambiguous state comparison \cite{barnett2003}, the error probability is limited to $\frac{1}{2}e^{-2\alpha^2}$ plotted by a green solid line \cite{ACJ06}.
Since an optimal measurement minimizing the error probability for binary coherent state comparison is accomplished by separately performing an optimal measurement for the BPSK states discrimination on each mode,
the Helstrom bound is reachable by introducing fast feedback operations to the displacement with the photon detection measurement \cite{Dolinar73,Becerra13,Becerra15,PhysRevApplied.13.054015}.

\section{Conclusions}
In this paper we have studied the state discrimination problem for quantum Gaussian states of light. 
For constant-$\hat{p}$ sets of states we have determined the optimal Gaussian discrimination protocol.
We found that the lower bound of the error probability for Gaussian strategies can be obtained by simply performing an $\hat{x}$-quadrature homodyne detection on each mode of the given state.
Although such sets of states as defined in Sec.~\ref{sec:gdiscriminate} may seem artificial, they cover many physically and technologically relevant problems.
As one of the relevant and important examples, we investigated the state comparison problem and, by applying the above statement, revealed that homodyne detections separately implemented on each mode is the best Gaussian measurement minimizing the error probability.
Moreover, we have discussed the performance for binary coherent state comparison with non-Gaussian strategies based on photon detections and compared them with the ultimate Gaussian limit.

\acknowledgements{WR and MT acknowledge the support of JST CREST Grant
No. JPMJCR1772.}


\input{main2.bbl}

\end{document}

%% file: main2.bbl
%